# Reflection and Transmission of Airy Pulse from Controllable Periodic Temporal Boundary

*Deependra Singh Gaur[1,*] and Akhilesh Kumar Mishra[1,2,#]*

[1]*Department of Physics, Indian Institute of Technology Roorkee, Roorkee- 247667, Uttarakhand, India*
[2]*Centre for Photonics and Quantum Communication Technology, Indian Institute of Technology Roorkee, Roorkee- 247667, Uttarakhand, India*

[#]akhilesh.mishra@ph.iitr.ac.in, [*]ds_gaur@ph.iitr.ac.in

## Abstract

We numerically investigate the interaction between two Airy pulses propagating at different wavelengths. The periodically varying peak intensity of the soliton that emerges from stronger Airy pulse (pump pulse) leads to the formation of periodic temporal boundary. The relatively weaker Airy pulse (probe pulse) on interaction with this boundary gets partially reflected as well as transmitted. As a result, the probe pulse spectrum splits into two parts- the reflected pulse spectrum undergoes red shift while transmitted pulse exhibits blue shift. The probe pulse witneses maximum reflection when point of interaction lies on the intensity maxima of the emergent soliton from pump Airy pulse. On the other hand, maximum transmission occurs when probe Airy pulse interacts at the intensity minima of the soliton. The reflection and transmission processes can be manipulated by tuning the time delay between pump and probe Airy pulses. In case of sufficiently intense pump pulse, the temporal boundary mimics the artificial optical event horizon, and the weak probe Airy pulse is completely reflected. This phenomenon is equivalent to the temporal version of total internal reflection. The time delay and truncation parameters play a vital role in the interaction. The results of the study hold potential applications in optical manipulation and temporal waveguiding.

**Introduction**

Spatial and temporal boundaries have distinct physical meaning. The spatial boundary breaks the translation symmetry, and the medium properties are different on either side of the interface [1]. An optical beam encountering the spatial boundary undergoes reflection and refraction which are explicated by Snell's law and Fresnel's coefficients [2]. On the other hand, consider a space of uniform refractive index at any instant of time whose refractive index changes uniformly in whole space with time. Using the space time duality, one may also ask the behaviour of optical pulses at a temporal boundary. The reflection and refraction of electromagnetic waves at a temporal boundary have also been reported where two physical processes named as impedance mismatch and time scaling due to sudden change in the speed of light, are involved [3]. The concept of time reflection and refraction in a nondispersive medium was described for the first time in the context of plasma [4].

Plansinis et. al. has reported temporal equivalent of reflection and refraction of optical beam where frequency is considered to be the time analogue of angle in space. If any spectral component of the incident pulse does not satisfy the phase matching or momentum conservation condition, then the incident pulse is totally reflected from the temporal boundary, which is temporal analogue of total internal reflection [5]. Analytical expressions have been derived for finding the reflection and transmission coefficients at a temporal boundary inside a dispersive media [6]. Presence of zero nonlinearity wavelength and the rise time of the temporal boundary significantly influence the reflection and refraction of the pulse at temporal boundary [7,8]. Reflection spectra retain their shape, provided the incident pulse is not chirped, whereas shape of transmission spectra is not influenced by the initial chirp parameter [9]. A recent report has discussed the energy conservation rules at the temporal boundary [10].

The event horizon is created in moving media when the local speed of a medium becomes larger than the speed of the wave moving in that medium. Cross phase modulation (XPM), a nonlinear phenomenon, is reported to create such temporal boundaries that can be utilized in confining the light pulses [11]. This led to the proposal of the concept of temporal waveguides [12,13]. Based on the XPM, fiber optical analogue of the event horizon is explored where soliton acts as temporal boundary [14]. In an optical fiber, realization of an artificial event horizon relies on the XPM between two co-propagating pulses centered at two different wavelengths [15]. An intense soliton modifies the effective refractive index of the optical fiber and this change in refractive index moves with the soliton pulse [16]. Effective properties of medium therefore gets modified while the medium itself remains at rest. On the other hand, probe pulse launched at different group velocity follows the soliton pulse and at a point of interaction group velocity of the probe pulse becomes smaller than the group velocity of the soliton pulse due to the change in refractive index. Owing to the change in group velocity, control pulse gets reflected and a frequency shift in the spectrum is observed. This not only introduces frequency shift in spectrum but also alters the group velocity of the soliton, which in turn leads to the acceleration in temporal propagation of the soliton [17]. It has also been reported that the optical event horizon is realised by soliton as well as its own dispersive wave, which results in new frequency generation [18-24].

Light manipulation ability of Airy pulse has been admired over the Gaussian pulse [25]. The reflection and refraction of the Airy pulse at sharp temporal boundary is also explored in ref. [26]. In general, strong *sech* pulse is employed as pump pulse in mimicking the artificial optical event horizon. Recent investigation unveiled that soliton from Airy pulse can also form an optical analogue of event horizon by creating the temporal boundary [27-30].

In this work, we examine the interaction of two Airy pulses, named strong pump and weak probe pulse. These pulses are assumed to propagate at different center wavelengths. Strong Airy pulse lies in anomalous dispersion regime therefore soliton shedding is supported, which acts as a temporal boundary for weak Airy pulse that is selected to propagate in normal dispersion regime. The transmission and reflection of probe Airy pulse are explored in different scenario of the amplitude and time delay variations. The results of study of the time delay variation hold a particular importance due to inherent periodicity in the temporal boundary.

**Theoretical Model**

We assume that the probe pulse propagates in a medium with the propagation constant $\beta(\omega)$. The propagation constant can be expanded in terms of Taylor series around its central frequency $\omega_0$,

$$\beta(\omega) = \beta_0 + \beta_1(\omega - \omega_0) + \frac{\beta_2}{2}(\omega - \omega_0)^2 \qquad (1)$$

where $\beta_1$ and $\beta_2$ are known as the inverse of the group velocity $v_{g2}$ and group velocity dispersion of the pulse respectively. If a co-propagating pump pulse moving with group velocity $v_{g1}$ creates a moving temporal boundary by inducing the nonlinear refractive index, then the dispersion relation in the moving frame becomes [5],

$$\beta'(\omega) = \beta_0 + \Delta\beta_1(\omega - \omega_0) + \frac{\beta_2}{2}(\omega - \omega_0)^2 + \beta_B(z)H(t - \tau_s), \qquad (2)$$

where $\Delta\beta_1 = \beta_1 - \frac{1}{v_{g1}} = \left(\frac{1}{v_{g2}} - \frac{1}{v_{g1}}\right)$ and $\beta_B(z) = \omega_0 \Delta n(z)/c$ is the change in propagation constant caused by nonlinear index change $\Delta n$ for $t > \tau_s$ which depends on the propagation distance $z$. Here $\tau_s$ is the time delay between the pulses. We would like to notice that $\Delta n$ is a function of position because of periodic nature of the soliton intensity profile. We assume that temporal boundary created by the pump pulse is sharp at $t = \tau_s$. Hence, the Heaviside function $H(t - \tau_s)$ takes the value 0 for $t < \tau_s$ and 1 for $t > \tau_s$.

With these assumptions, we apply the momentum conservation on $\beta'$ and set $\beta'(\omega) = \beta_0$, in eqn. (2) that results,

$$\frac{\beta_2}{2}(\omega - \omega_0)^2 + \Delta\beta_1(\omega - \omega_0) + \beta_B(z)H(t - \tau_s) = 0. \qquad (3)$$

For $t < \tau_s$, last term will be vanished from eqn. (3) and we obtain following two solutions,

$$\omega_i = \omega_0, \qquad \omega_r = \omega_0 - 2\left(\frac{\Delta\beta_1}{\beta_2}\right). \qquad (4)$$

These solutions respectively represent incident ($\omega_i$) and reflected ($\omega_r$) frequencies.

For $t > \tau_c$, last term on LHS in eqn. (3) has a finite value and sudden change in refractive index introduces position dependent variable propagation constant $\beta_B(z)$ that changes periodically with the propagation distance for Airy soliton. In this case, eqn. (3) provides the solution for transmitted frequency as

$$\omega_t = \omega_i + \frac{\Delta\beta_1}{\beta_2}\left[-1 \pm \sqrt{1 - \frac{2\beta_B(z)\beta_2}{(\Delta\beta_1)^2}}\right]. \qquad (5)$$

Eqn. (5) can be considered as a valid physical solution only for positive sign [5]. In addition, if $\Delta\beta_1 \gg \sqrt{\beta_B\beta_2}$, then eqn. (5) can be approximated as

$$\omega_t = \omega_i - \frac{\beta_B(z)}{\Delta\beta_1} = \frac{k_0 \Delta n}{\Delta\beta_1} \qquad (6)$$

Frequency shift in transmitted pulse spectrum depends on the sign of $\beta_B$.

These results can be verified numerically by considering two pulse interaction through the XPM where strong pump pulse creates moving temporal boundary, and the reflection and transmission of the dispersive probe pulse are studied. Hence, we consider two light pulses propagating in a single mode fiber at different wavelengths. The pump ($A_1$) and probe ($A_2$) pulse envelopes dynamics are governed by the coupled nonlinear Schrödinger equations (CNSEs). The dimensionless form of CNSEs are expressed as [25,31,32]

$$\frac{\partial A_1}{\partial z} + \frac{i}{2} sgn(\beta_{21}) \frac{\partial^2 A_1}{\partial \tau^2} = iN_1[|A_1|^2 + 2|A_2|^2]A_1, \qquad (7)$$

$$\frac{\partial A_2}{\partial z} + \psi \frac{\partial A_2}{\partial \tau} + \frac{i}{2} \delta_{22} \frac{\partial^2 A_2}{\partial \tau^2} = iN_2[|A_2|^2 + 2|A_1|^2]A_2, \qquad (8)$$

where $z = Z/L_{D1}$ and $\tau = T/\tau_0$ are the normalized propagation distance and normalized time in pulse frame respectively. In above equations, the parameter $\psi$ is defined as $L_{D1}/L_W$ where

$L_{D1} = \tau_0^2/|\beta_{21}|$ and $L_W = \left(\frac{v_{g2}^{-1}-v_{g1}^{-1}}{\tau_0}\right)$ respectively. The parameter $sgn(\beta_{21})$ denotes the sign of the group velocity dispersion (GVD) experienced by pump pulse while $\delta_{22} = \beta_{22}/|\beta_{21}|$, where $\beta_{22}$ represent the GVD coefficient of the probe pulse. The nonlinear parameters are defined as $N_1 = \gamma_{11}P_0\tau_0^2/|\beta_{21}|$ and $N_2 = N_1\frac{\omega_2}{\omega_1}$ where $\gamma_{11} = n_2\omega_1/cA_{eff}$. The initial pulse envelopes of pump and probe pulses are expressed respectively as

$$A_1 = f(a_1)Ai(\tau - \tau_s)\exp(a_1(\tau - \tau_s)), \qquad (9)$$

and
$$A_2 = rf(a_2)Ai(\tau - \tau_c)\exp(a_2(\tau - \tau_c)), \qquad (10)$$

where $r$ denotes the amplitude ratio of the probe and pump pulses while $f(a_1)$ and $f(a_2)$ are numerically calculated parameters which keep the peak value of Airy function at 1. Parameters $\tau_s$ and $\tau_c$ are the pump and probe pulse time delays respectively and $a_1$ and $a_2$ are the corresponding truncation parameters. For simulation the pump and probe pulse frequencies are chosen to be $\omega_s = 0.6\ PHz$ and $\omega_c = 1.8\ PHz$ respectively and pulse width for both pulses are taken to be $\tau_0 = 21$ fs [16]. These parameters are chosen in such a way that the pump and probe pulses experience anomalous and normal dispersions respectively. The numerical values of the dispersion and nonlinear parameters are taken as $\beta_{21} = -0.229$ fs²/µm, $\gamma_{11} = 0.1$ W⁻¹m⁻¹, $\beta_{22} = 0.08$ fs²/µm and $\gamma_{12} = 0.3$ W⁻¹m⁻¹ [16]. In order to solve eqn. (7) and (8), we use the following normalized parameter - $sgn(\beta_{21}) = -1$, $\delta_{22} = 0.35$, $N_1 = 1$, $N_2 = 3$ and $\psi = 1$.

## Results and Discussion

Here, we study the two Airy pulse interaction mediated through XPM in a dispersive and Kerr nonlinear medium. The strong pump Airy pulse propagates in anomalous dispersion regime whereas the weak Airy pulse propagates in normal dispersion regime. Probe Airy pulse is launched prior to the pump Airy pulse with slightly smaller group velocity than the pump Airy pulse. The values of truncation parameters are chosen as $a_1 = a_2 = 0.25$ unless mentioned otherwise.

### 1. Effect of the amplitude of the probe Airy pulse

Equation (7) and (8) are numerically solved simultaneously to model the dynamical evolution of the pulses. The temporal evolutions of probe and pump Airy pulses with propagation distance are shown in fig. 1 (a). Soliton emerges from the strong pump Airy pulse whose intensity varies periodically with propagation distance [27]. Emergent soliton introduces a nonlinearity induced co-propagating refractive index change, which acts as a temporal boundary for probe Airy pulse. Weak probe Airy pulse collides with the emergent soliton and subsequently undergoes reflection due to the nonlinear refractive index change. Group velocities of both pulses are increased after the collision and the pulses suffer shift towards the leading edge [16]. In addition to the reflection through this temporal boundary, depending upon the strength of temporal boundary and amplitude of the incident probe Airy pulse, the weak Airy pulse undergoes refraction (transmission). Temporal boundary induced variable propagation constant $\beta_B(z)$ significantly affect the transmitted pulse spectrum as expressed in eqn. (6). Here, $\beta_B(z)$ is a function of propagation distance. If $\beta_B(z)$, in eqn. (6), assumes such a value that gives $\omega_t = 0$, then the transmission will not be possible, and the incident pulse will suffer complete reflection- a phenomenon analogous to total internal reflection [5]. Owing to the change in group velocity, pump pulse spectrum shifts towards the blue side as shown in fig. 1 (b). Importantly, the probe Airy pulse spectrum splits into two parts that corresponds to reflection and refraction. A large red shift in frequency is observed in the probe pulse spectrum that corresponds to the reflected pulse as delineated in fig. 1 (c) and this redshift agrees with the prediction of eqn. (4). Moreover, a narrow blue band is observed which corresponds to the transmitted pulse spectrum. An optical pulse propagating in anomalous dispersion regime creates quantum mechanical equivalent of repulsive potential that causes $\beta_B$ to be negative [19]. Thus, according to eqn. (6), a blue shift is observed in the transmitted spectrum.

The spectral and temporal dynamics of the pump and probe pulses are significantly modified by the amplitude ratio $r$. Fig. 2 (a) denotes the spectral profile of the emergent pump soliton at output ($z = 80$) for different peak amplitude of probe Airy pulse. Pump pulse spectrum shifts more towards blue as $r$ increases. Besides, the temporal shift of the soliton is also increased as depicted in fig. 2(b). While the peak intensity of the emergent soliton and its periodicity remains unaffected due to the change in amplitude ratio $r$ (see fig. 2(c)).

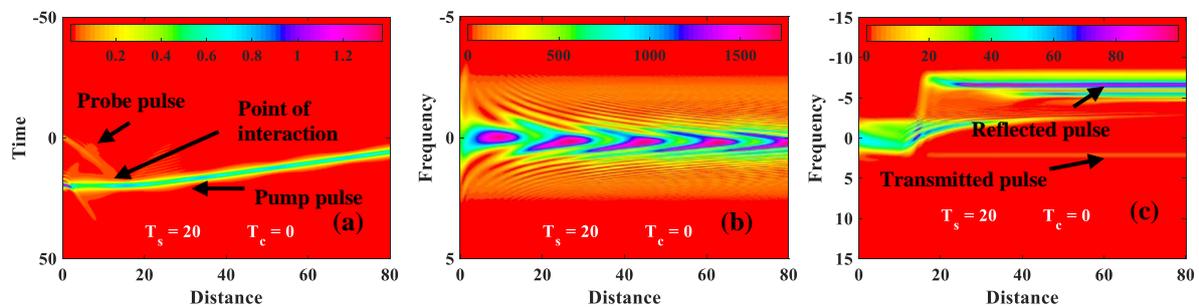

Fig. 1 (a) Temporal evolution of pump and probe Airy pulses, (b) spectral evolution of pump Airy pulse and (c) spectral evolution of probe Airy pulse for $r = 0.30$.

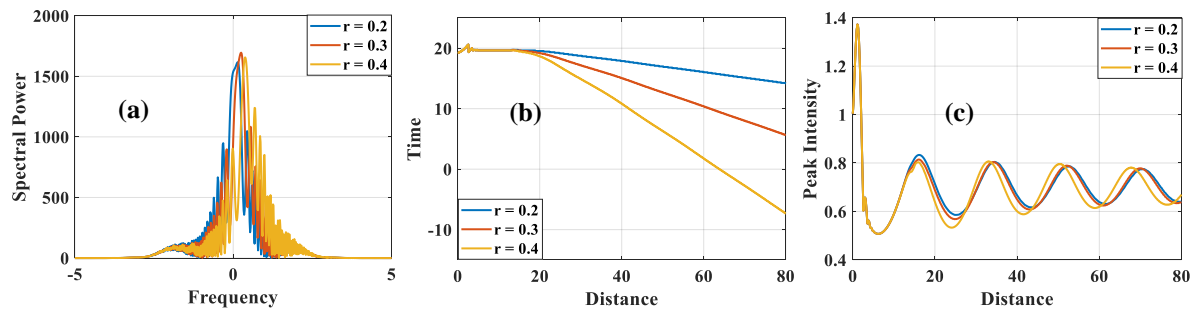

Fig. 2 (a) Output spectrum (b) temporal position and (c) peak intensity of pump Airy soliton for different peak amplitudes of probe Airy pulse.

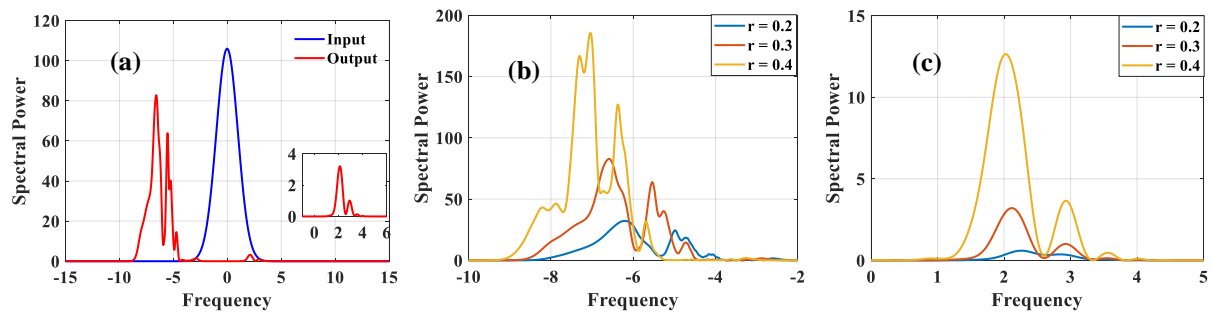

Fig. 3 (a) Input and output spectra at $r = 0.30$, (b) reflected spectra and (c) transmitted spectra of probe Airy pulse for different peak amplitude of probe pulse. Inset in fig. 3 (a) shows the spectrum of the blue shifted transmitted pulse.

Fig. 3 (a) shows the spectral profiles of the probe Airy pulse at the input as well as at the output ($z = 80$) where the amplitude ratio $r = 0.30$. Although a large red spectral shift is observed but a small part of the spectrum is found blue shift. The red shifted spectrum corresponds to the reflected probe pulse whereas the blue shifted part is the signature of the transmitted part. The inset in fig. 3 (a) shows the clear spectral profile of the blue shifted transmitted pulse. Reflected and transmitted pulse spectra for different amplitude ratio are delineated respectively in figs. 3 (b) and (c) which reveal that spectral power of both transmitted and reflected pulses increase with increasing $r$. The shape of the transmitted pulse spectrum is Airy but the reflected pulse spectrum gets distorted and multipeak structure is observed. Airy spectrum of transmitted pulse shows that pulse acquires a quadratic temporal phase during the collision process.

## 2. Effect of time delay on reflection and transmission

The time delay between the pump and probe pulses significantly modifies the collision dynamics. In this section, we explore the impact of time delay on the spectral and temporal evolution of the pump and probe pulses. As stated before, the periodic intensity modulation of the emergent soliton induces a periodic nonlinear refractive index which appears as a periodic moving temporal boundary. Therefore, the strength of the temporal boundary becomes position dependent. Hence different time delays alter the collision position. Fig. 4 (a1) shows the collision between probe and pump pulses at a point where intensity of the pump soliton is maximum. Owing to the change in group velocity, emergent soliton shifts towards the leading edge. The probe pulse is reflected completely at this point of interaction (which can be consider as temporal equivalent of total internal reflection) as shown in fig. 4 (a2). Moreover, reflected probe pulse spectrum is observed to be red shifted as delineated in fig. 4 (a3).

In second case, we change the time delay between pump and probe pulses such that interaction occurs at the position where emergent soliton intensity drops to a local minimum as depicted in fig. 4 (b1). Therefore, probe Airy pulse reflects as well as transmits through this relatively weaker temporal boundary and spectrum of probe pulse splits into two parts (see fig. 4 (b2)). The reflected pulse spectrum gets red shifted while blue shifted spectrum is observed for transmitted pulse as shown in fig. 4 (b3).

In order to show periodic total internal reflection, we again vary the time delay in such a way that interaction happens at successively next maximum of intensity as shown in fig. 4 (c1). The probe pulse reflection from temporal boundary can be seen in fig. 4 (c2). We observe red shift in probe pulse spectrum and the blue shifted transmitted spectra is again fades away as illustrated in fig. 4 (c3). Further, the interaction is explored at the position of successively next intensity minimum by increasing time delay between pulses as depicted in fig. 4 (d1). The probe Airy pulse undergoes reflection and transmission both, but the temporal intensity of the transmitted pulse is weak, therefore it seems imperceptible in fig 4 (d2). Although, transmitted pulse is not visible in time domain but the corresponding frequency spectrum at blue region is present as shown fig. 4 (d3). The numerical values of the time delays are mentioned in all the figures. This investigation manifests that emergent soliton from Airy pulse forms a periodic temporal boundary by inducing the periodic nonlinear refractive index. Therefore, periodic total internal reflection from periodic temporal boundary is observed.

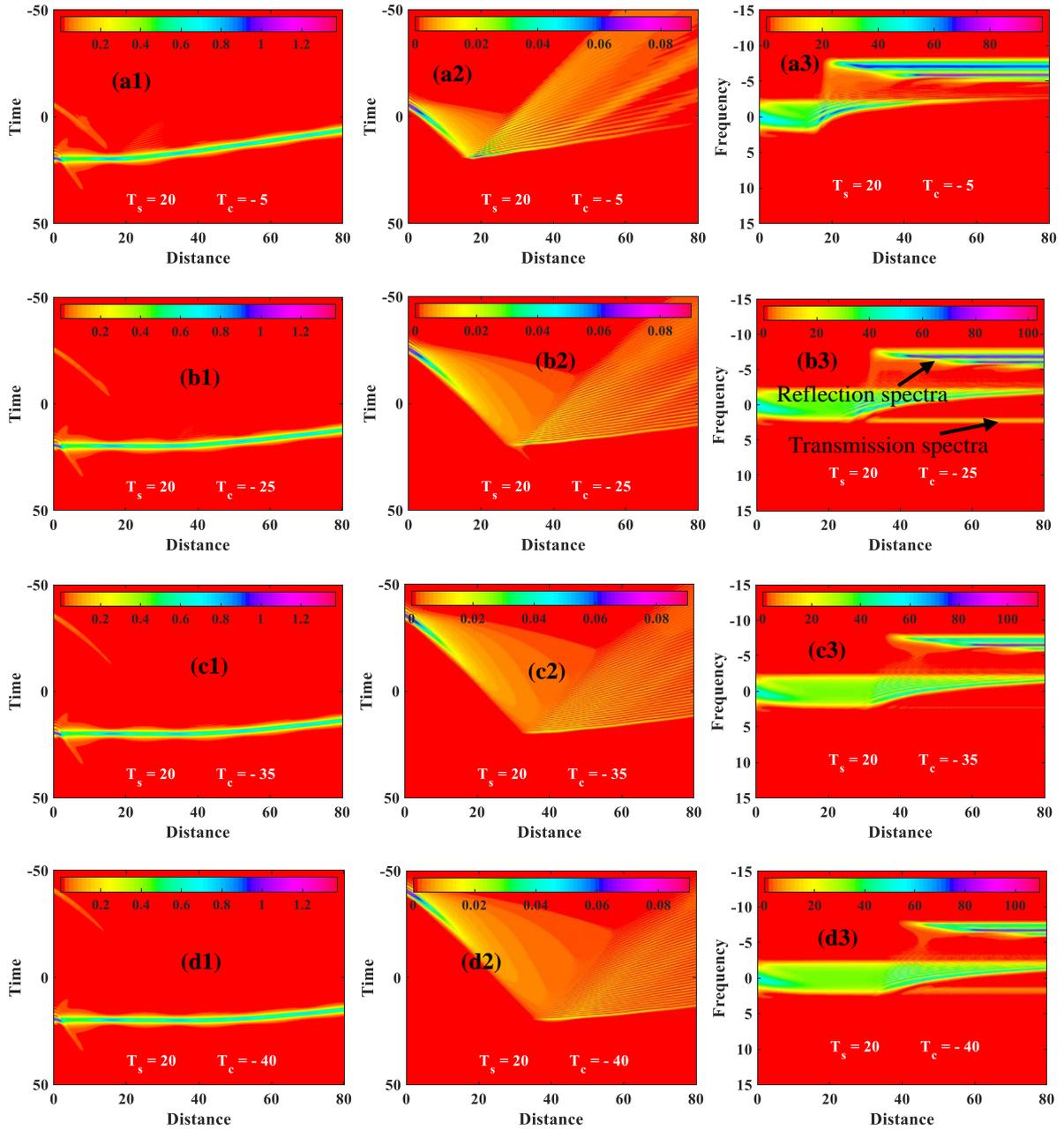

Fig. 4 Temporal evolution of pump (first column), probe pulse (second column) and spectral evolution of probe (third column) pulse for different values of $T_c$.

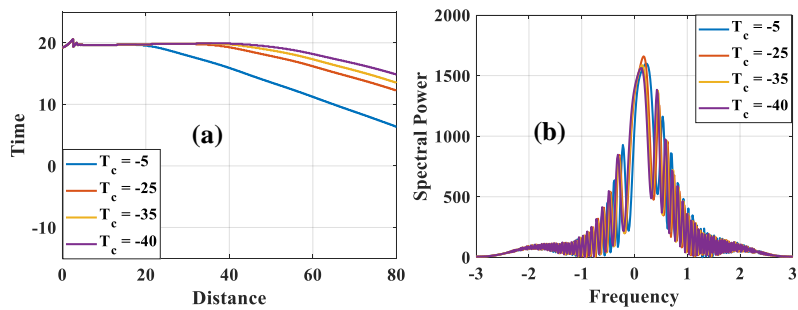

Fig. 5 (a) Time shift and (b) spectrum of pump pulse at output for $T_s = 20$ and different value of $T_c$

Temporal shift with propagation distance and spectral profiles of the emergent pump soliton at the output for different time delays are illustrated respectively in fig. 5 (a) and (b) which manifest that interaction is delayed for large value of $T_c$ therefore small temporal shift is observed. As the time delay is reduced, the pulse interaction becomes stronger that results in large temporal shift as shown in fig. 5 (a). On the other hand, pulse spectrum of the emergent pump soliton remains almost unchanged as illustrated in fig. 5 (b).

## 3. Effect of truncation parameter

Truncation parameter modifies the propagation dynamics considerably. In the present simulation, the truncation parameter of pump and probe Airy pulses are taken respectively to be $a_1 = 0.15$ and $a_2 = 0.25$ and the interaction is shown in fig. 6 (a). The temporal shift of the emergent soliton is reduced as compared to the case $a_1 = 0.25$ (compare fig. 6 (a) and fig. 1 (a)). The probe Airy pulse suffers both reflection as well as transmission through this boundary and the amplitude of the transmitted pulse is enhanced significantly as shown in fig. 6 (b). Consequently, the spectrum corresponding to transmitted pulse is pronounced as depicted in fig. 6 (c).

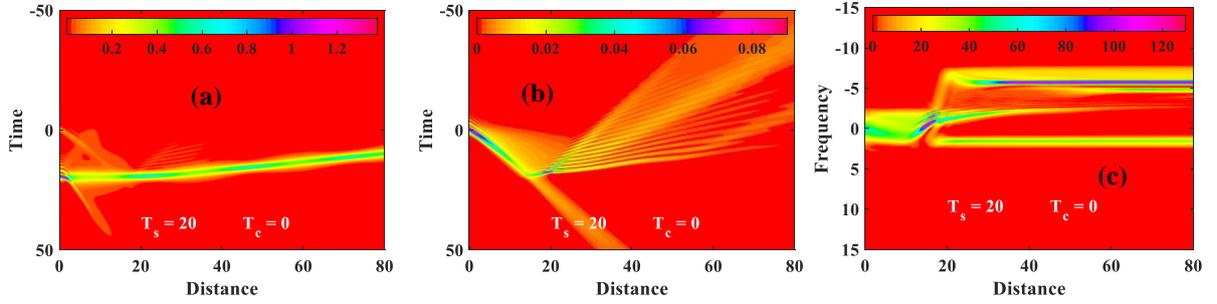

Fig. 6 (a) Temporal evolution of pump and probe pulses (b) temporal evolution of probe Airy pulse and (c) spectral evolution of probe Airy pulse for $a_1 = 0.15$ and $a_2 = 0.25$

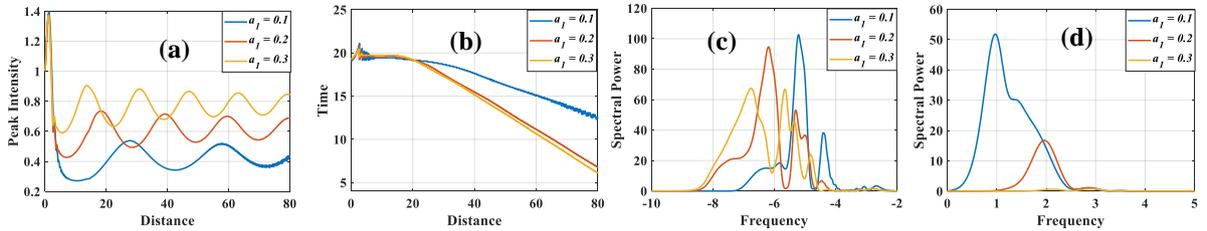

Fig 7. (a) Peak intensity (b) temporal position of the pump Airy pulse (c) reflected and (d) transmitted probe pulse spectrum for $a_2 = 0.25$ and different value of $a_1$.

The truncation parameter of the pump Airy pulse not only alters the maximum intensity of the emergent soliton but also changes its periodicity as evident from fig. 7 (a). Owing to this fact, the temporal shift of emergent soliton, reflection and transmission of probe Airy pulse change significantly. A small value of truncation parameter causes a less intense soliton shedding which leads to the formation of temporal boundary of weaker strength. Less intense soliton induces weak nonlinear refractive index change therefore the amplitude of transmitted pulse is increased as shown in fig. 6 (b). In addition, collision of this weak emergent soliton and probe Airy pulse results in small increase in group velocity consequently, temporal shift of emergent soliton is reduced. Temporal shift of the emergent soliton is increased as the value of truncation parameter is increased as shown in fig. 7 (b). The reflected and transmitted spectrum is depicted in fig. 7 (c) and (d) respectively.

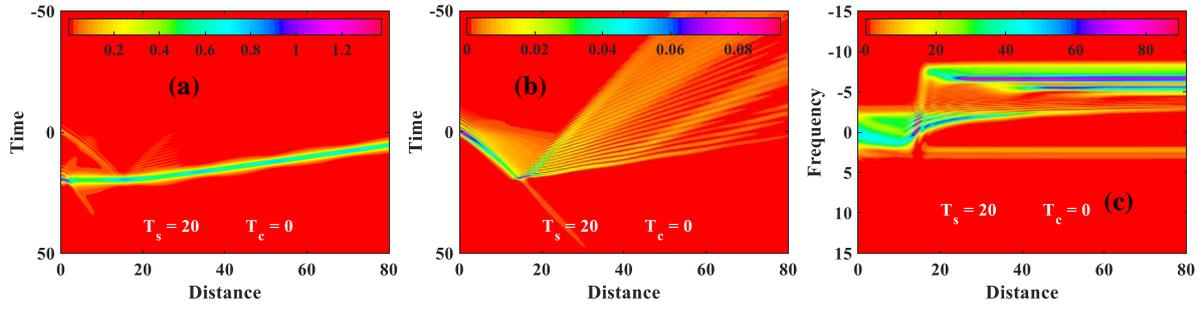

Fig. 8 (a) Temporal evolution of pump and probe Airy pulse, (b) temporal evolution of probe Airy pulse and (c) spectral evolution of probe Airy pulse with truncation $a_1 = 0.25$ and $a_2 = 0.15$

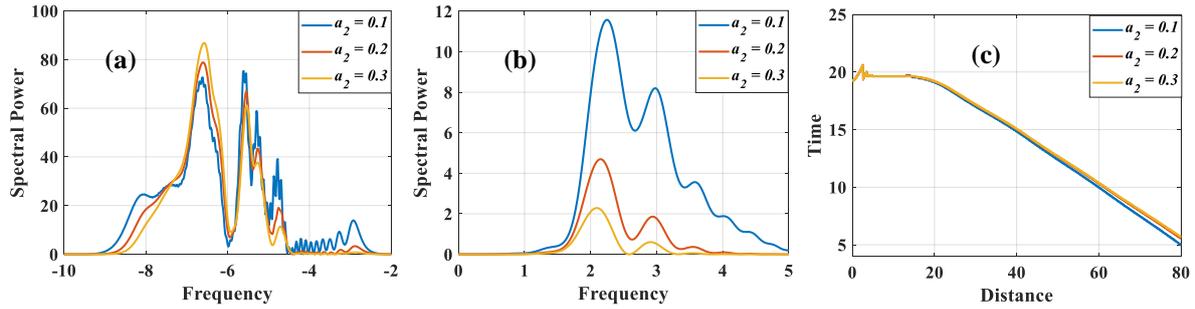

Fig. 9 (a) Reflected (b) transmitted probe pulse spectrum and (c) temporal position of pump pulse for $a_1 = 0.25$ and different value of $a_2$

Next, we vary the truncation parameter of probe pulse. Here we chose $a_1 = 0.25$ and $a_2 = 0.15$ in simulation. The temporal evolution of the interaction is shown in fig. 8 (a) where the emergent soliton is shifted towards the leading edge. For small values of the truncation parameter, the Airy tail extends longer. The transmitted probe Airy is shown in fig. 8 (b). The corresponding spectrum is shown in fig. 8 (c).

Fig. 9 (a) and (b) show the spectral profile of transmitted and reflected pulse at output for different values of truncation parameter. The reflected pulse shape is found distorted with multi red shifted peaks. As the value of truncation is reduced, Airy shape structure appears at blue side in the reflected pulse spectrum appears because frequency shift is delayed as depicted in fig.9 (a). The transmitted pulse shape remains asymmetric Airy whose amplitude increases with decreasing truncation parameter as shown in fig. 9 (b). The truncation parameter of probe Airy pulse does not influence the temporal shift of the emergent pump soliton as clearly illustrated in fig. 9 (c).

**Conclusion**

In summary, we have shown the interaction of two Airy pulses propagating in opposite dispersion regimes. The pump Airy pulse propagating in anomalous dispersion regime exhibited soliton shedding. The soliton induced a nonlinear refractive index in the medium that consequently led to the formation of a moving temporal boundary. Since the intensity of the emergent soliton varies periodically with propagation distance, the strength of the temporal boundary was also periodic. The relatively weak probe Airy pulse suffered reflection and transmission through this boundary and the spectrum of reflected probe pulse shifted towards red while the blue shift is observed for the transmitted pulse. Importantly, the probe pulse exhibited temporal equivalent of total internal reflection at the temporal boundary. In addition, the emergent pump soliton is shifted towards the leading edge and the amount of shift depends upon the amplitude ratio between the two pulses, time delay between the pulses and truncation parameter $a_1$ of the pump Airy pulse. These novel characteristics are expected to facilitate optical manipulation, temporal waveguides and all optical switching.

**Acknowledgement**

Deependra Singh Gaur wishes to IIT Roorkee and Ministry of Education (MoE) India for providing the fellowship assistance. The encouragement from Prof. G. P. Agrawal is fully acknowledged.

**References**

1. B. E. A. Saleh and M. C. Teich, Fundamentals of Photonics, 2nd ed. (Wiley-Interscience, Hoboken, NJ, 2007).

2. M. Born and E. Wolf, Principles of Optics, 7th ed. (Cambridge University, 1999).

3. J. T. Mendonca and P. K. Shukla, "Time refraction and time reflection: two basic concepts" Phys. Scrip 65, 160-163 (2002).

4. Y. Xiao, D. N. Maywar and G. P. Agrawal, "Reflection and transmission of electromagnetic waves at a temporal boundary" Opt. Lett. 39, 574-577 (2014).

5. B. W. Plansinis, W. R. Donaldson, and G. P. Agrawal, "What is the Temporal Analog of Reflection and Refraction of Optical Beams ? " Phys. Rev. Lett. 115 183901 (2015).

6. J. Zhang, W. R. Donaldson and G. P. Agrawal, "Temporal reflection and refraction of optical pulses inside a dispersive medium: an analytic approach" JOSA B 38, 997-1003 (2021).

7. J. Zhang, W. R. Donaldson and G. P. Agrawal, "Impact of the boundary's sharpness on temporal reflection in dispersive media" Opt. Lett. 46 4053-4056 (2021).

8. A. C. Sparapani, J. Bonetti, N. Linale, S. M. Hernandez, P. I. Fierens, and D. F. Grosz, "Temporal reflection and refraction in the presence of a zero-nonlinearity wavelength" Opt. Lett. 48 339-342 (2023).

9. W. Cai, Z. Yang, H. Wu, L. Wang, J. Zhang, and L. Zhang, "Effect of chirp on pulse reflection and refraction at a moving temporal boundary" Opt. Express 30 34875-34886 (2022).

10. K. B. Tan, H. M. Lu, and W. C. Zuo, "Energy conservation at an optical temporal boundary" Opt. Lett. 45, 6366-6369 (2020).

11. B. W. Plansinis, W. R. Donaldson and G. P. Agarwal, "Cross-phase-modulation-induced temporal reflection and waveguiding of optical pulses" JOSA B 35 436-445 (2018).

12. J. Zhou, G. Zheng and J. Wu, "Comprehensive study on the concept of temporal optical waveguide" Phys. Rev. A 93 063847 (2016).

13. B. W. Plansinis, W. R. Donaldson and G. P. Agarwal, "Temporal waveguides for optical pulses" JOSA B 33 1112-1119 (2016).

14. K. E. Webb, M. Erkintalo, Y. Xu, N. G. R. Broderick, J. M. Dudley, G. Genty, and S. G. Murdoch, "Nonlinear optics of fibre event horizons," Nat. Commun. 5, 4969 (2014).

15. T. G. Philbin, C. Kuklewicz, S. J. Robertson, S. Hill, F. König, and U. Leonhardt, "Fiber-optical analog of the event horizon," Science 319, 1367–1370 (2008).

16. A. Demircan, S. Amiranashvili, and G. Steinmeyer, "Controlling light by light with an optical event horizon," Phys. Rev. Lett. 106. 163901(2011).

17. J. Gu, H. R. Guo, S. F. Wang, and X. L. Zeng, "Probe-controlled soliton frequency shift in the regime of optical event horizon," Opt. Express 23(17), 22285-22290 (2015).

18. D. V. Skryabin and A. V. Yulin, "Theory of generation of new frequencies by mixing of solitons and dispersive waves in optical fibers," Phys. Rev. E 72(1), 016619 (2005).

19. J. Drori, Y. Rosenberg, D Bermudez, Y. Silberberg and U. Leonhardt, "Observation of stimulated hawking radiation in an optical analogue" Phys. Rev. Lett. 122 010404 (2019).

20 A. Choudhary and F. König, "Efficient frequency shifting of dispersive waves at solitons," Opt. Express 20(5), 5538-5546 (2012).

21. S. Robertson and U. Leonhardt, "Frequency shifting at fiber-optical event horizons: The effect of Raman deceleration" Phys. Rev A 81 063835 (2010).

22. A.V. Gorbach and D.V. Skryabin, "Bouncing of a dispersive pulse on an accelerating soliton and stepwise frequency conversion in optical fibers" Opt. Express 15 14560-14565 (2007).

23. S. F. Wang, A. Mussot, M. Conforti, A. Bendahmane, X. L. Zeng and A. Kudlinski, "Optical event horizons from the collision of a soliton and its own dispersive wave", Phys. Rev. A 92 023837 (2015).

24. A. Yang, Y. He, S. Wang, and X. Zeng, "Manipulating Airy pulse in the regime of optical event horizon" Opt. Express 26 34689-34698 (2016).

25. W. Cai, M. S. Mills, D. N. Christodoulides and S. Wen, "Soliton manipulation using Airy pulses" Opt. Comm. 127-131 (2014).

26. W. Cai, Y. Tian, L. Zhang, H. He, J. Zhao and J. Wang, "Reflection and refraction of an Airy pulse at a moving temporal boundary" Ann. Phys. 532 (10), 2000295 (2020).

27. D. S. Gaur, A. Purohit and A. K. Mishra, "Soliton shedding from Airy pulses in a highly dispersive and nonlinear medium", JOSA B, 38, 3729-3736 (2021).

28. A. Purohit, D. S. Gaur and A. K. Mishra, "Dynamics of a chirped Airy pulse in a dispersive medium with higher-order nonlinearity" JOSA B, 38, 3605-3615 (2021).

29. D. S. Gaur and A. K. Mishra, "Impact of harmonic potential induced nonlinearity on Airy pulse propagation" J. Opt. 24(6), 065504 (2022).

30. D. S. Gaur and A. K. Mishra, "Ballistic dynamics of emergent soliton from Airy pulse in a medium with linear optical potential" Opt. and Laser Tech. 168 109996 (2024).

31. G. P. Agrawal, P. L. Baldeck and R. R. Alfano, "Temporal and spectral effects of cross-phase modulation on copropagating ultrashort pulses in optical fibers" Phys. Rev. A 40 5063-5072 (1989).

32. D. S. Gaur and A. K. Mishra, "Manipulating the Strong Airy Pulse Utilizing Weak Airy Pulse in the Regime of Optical Event Horizon," in *Frontiers in Optics + Laser Science 2023 (FiO, LS)*, paper JM7A.45.